\RequirePackage{lineno}
\documentclass[prd,twocolumn,showpacs,amsmath,amssymb,nofootinbib]{revtex4-1}
\usepackage{epsfig}
\usepackage{dcolumn}
\usepackage{graphicx}
\usepackage{subfigure}
\usepackage{times}
\usepackage{setspace}
\usepackage{float}
\usepackage{color}
\usepackage{pstricks}
\usepackage{pst-node}
\usepackage{rotating}
\usepackage{overpic}
\usepackage{multirow}

\newcommand{\ppp}{\pi^+\pi^- \pi^0}

\newcommand{\pp}{\pi^+\pi^-}
\newcommand{\kk}{K^+K^-}
\newcommand{\ppb}{p\bar{p}}

\newcommand{\ppkk}{\pi^+\pi^-K^+K^-}

\newcommand{\bfg}{\begin{figure}}
\newcommand{\efg}{\end{figure}}
\newcommand{\bitm}{\begin{itemize}}
\newcommand{\eitm}{\end{itemize}}
\newcommand{\bnum}{\begin{enumerate}}
\newcommand{\enum}{\end{enumerate}}
\newcommand{\btbl}{\begin{table}}
\newcommand{\etbl}{\end{table}}
\newcommand{\btbu}{\begin{tabular}}
\newcommand{\etbu}{\end{tabular}}
\newcommand{\bcl}{\begin{center}}
\newcommand{\ecl}{\end{center}}
\newcommand{\bbt}{\bibitem}
\newcommand{\beq}{\begin{equation}}
\newcommand{\eeq}{\end{equation}}
\newcommand{\beqr}{\begin{eqnarray}}
\newcommand{\eeqr}{\end{eqnarray}}

\begin{document}
\normalsize
\parskip=5pt plus 1pt minus 1pt

\title{\boldmath {Observation of the first iso-spin Charmonium-like State $Z_c(4020)$ }}

\author{
  \begin{small}
    \begin{center}
      JI Qing-ping$^{1}$, GUO Yu-Ping$^{2}$, GUO Ai-Qiang$^{3}$,\\
      YU Chun-Xu$^{1}$, WANG Zhi-Yong$^{3}$
      \\
      \vspace{0.2cm}
      \vspace{0.2cm} {\it
        $^{1}$Department of Physics, Nankai University, Tianjin 300071, People's Republic of China\\
$^{2}$ Johannes Gutenberg University of Mainz, Johann-Joachim-Becher-Weg 45, D-55099 Mainz, Germany\\
$^{3}$ Institute of High Energy Physics, Beijing 100049, People's Republic of China\\
      }\end{center}
    \vspace{0.4cm}
  \end{small}
}

\affiliation{}

\begin{abstract}
In this paper, we present a new experimental progress in brief on the recent observation of the charged charmonium-like state $Z_{c}(4020)^{\pm}$  states and its iso-spin partner $Z_{c}(4020)^{0}$ in the $e^+e^¨C \to \pi^{+,0}\pi^{-,0} h_{c}$ process of at the BESIII experiment. The charged $Z_{c}(4020)$ is its decay into $\pi^{\pm} h_{c}$ final state, and carries electric charge, thus it contains at least four quarks. The observation of both charge and neutral state makes $Z_{c}(4020)$ the first iso-spin triplet $Z_{c}$ state observed in experiment.
\end{abstract}

\maketitle
\section{Introduction} \label{sec::introduction}
In quark model, hadrons are made up of quarks. At the lowest configuration, one quark and one anti-quark form meson while three quarks do baryon~\cite{lab1}. But hadrons with other configurations, called exotic states, are not excluded, such as deuteron, which is formed with a proton and a neutron~\cite{lab2,lab3}. The search for these exotic states has stimulated great interest. Although no convincible evidence has been found in experimental, strong evidence for mesons that do not fit into the simple qq-bar scheme of the quark model has been accumulating during the past decade, such as X(3872)~\cite{lab4,lab5,lab6,lab7}, Y(4260)~\cite{lab8}, $Z(4430)^{+}$~\cite{lab9,lab10,lab11,lab12} in charmonium region.

Experimentally, what important to establish exotic states is finding clear different signatures compared with the conventional hadrons. One promising method is to search for electrically charged states that can decay into hidden charm final states, since such states contain at least four quarks, and therefore can not be conventional mesons, if they are mesons.

In charmonium region, the first such state reported in experiment is $Z(4430)^{+}$ from Belle in 2008 in a study of $B\to K\pi^{+}\psi(2S)$ decays~\cite{lab9}. This state has not been confirmed from BaBar~\cite{lab11} using the same analysis method. Until the result reported by LHCb in 2014~\cite{lab12}, the $Z(4430)^{+}$ has been confirmed with ambitious significance, and the quantum number is determined to be $1^{+}$.
The first confirmed charged charmoniumlike state is $Z_{c}(3900)$ observed in $\pi^{\pm}J/\psi$ mass spectrum in the $e^{+}e^{-}\to\pi^{+}\pi^{-}J/\psi$
process at BESIII using data sample taken at $\sqrt{s}=4.26$ GeV~\cite{lab13}. This state has been confirmed shortly by Belle experiment~\cite{lab14}
and CLEO-c~\cite{lab15} using same decay mode. Besides, in CLEO-c data, an evidence of neutral $Z_{c}(3900)$ has also been reported in the neutral process. The observation of $Z_{c}(3900)$ has attracted great interest from both theoretical and experimental side. Different models have been proposed, such as tetraquark consisting of $cu$ and $\bar{c}\bar{d}$ diquarks~\cite{lab16}, molecule consisting of $D^{*}$ and $D$~\cite{lab17}, hadroquarkonium consisting of $ud$ bound to a color-singlet $c\bar{c}$ core\cite{lab18}, or other configurations~\cite{lab19}. Almost in the same, the study of the $(DD^{*})^{\pm}$ and $(D^{*}D^{*})^{\pm}$ system in $(DD^{*})^{\pm}\pi$ and $(D^{*}D^{*})^{\pm}\pi$ states in the same data sample have been performed, and strong near-threshold peaks has been observed, which is named as $Z_{c}(3885)$ \cite{lab20} and $Z_{c}(4025)$~\cite{lab23}, respectively.

Apparently, searching for similar states in the invariant mass spectrum made up of one pion and other charmonium states such as $h_{c}$, $\chi_{cJ}$, and so on is interesting. Any observation may provide more information on understanding these states, and helpful in building relationship between those states. Such studies have been performed at BESIII in both $e^{+}e^{-}\to\pi^{+}\pi^{-}h_{c}$ and $\pi^{0}\pi^{0}h_{c}$ processes~\cite{lab21,lab22}.
A distinctive structure, named as $Z_{c}(4020)$ has been observed in the invariant mass spectrum of $\pi h_{c}$. The mass of $Z_{c}(4020)$ is close to that of the mentioned $Z_{c}(4025)^{\pm}$ observed in $(D^{*}D^{*})^{\pm}\pi$ states. Here we just introduce the observation of both electrically charged and neutral $Z_{c}(4020)$ at BESIII.

\begin{figure}[!htbp]
\begin{center}
\includegraphics[width=0.45\textwidth]{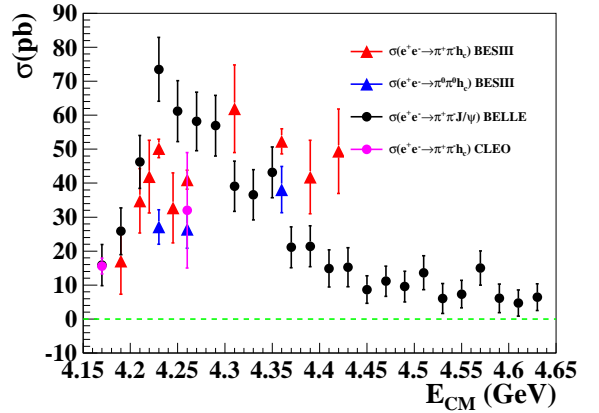}
\caption{ The comparison between $\sigma(e^{+}e^{-}\to\pi^{+}\pi^{-}h_{c})$ and $\sigma(e^{+}e^{-}\to\pi^{+}\pi^{-}J/\psi)$. The errors are statistical only.}
\label{fig fig1}
\end{center}
\end{figure}

\begin{table*}[!htbp]
\begin{center}
\renewcommand{\arraystretch}{1.2}
\caption{ The cross section of $e^{+}e^{-}\to\pi\pi h_{c}$ at different energy points and the ratios ($R_{\pi\pi h_{c}}$)between charged and neutral modes. For the first three energy points, besides the upper limits, the central values and the statistical errors are also listed. The second errors are systematic errors and the third ones are from the measurement of branching fraction of $h_{c}\to\gamma\eta_{c}$~\cite{lab26}.} \label{tab::tab1}
\begin{tabular}{c c c c }
\hline \hline
$\sqrt{S}$(GeV)&$\sigma(e^{+}e^{-}\to\pi^{+}\pi^{-}h_{c})$ &$\sigma(e^{+}e^{-}\to\pi^{0}\pi^{0}h_{c})$& $R_{\pi\pi h_{c}}$\\\hline
3.900&$0.0\pm6.0$ or $<$8.3 &-&-\\
4.009&$1.9\pm1.9$ or $<$5.0 &-&- \\
4.090&$0.0\pm7.4$ or $<$13 &-&-\\
4.190&$17.7\pm9.8\pm1.6\pm2.8$ &-&- \\
4.190&$17.7\pm9.8\pm1.6\pm2.8$  &-&-\\
4.210&$34.8\pm9.5\pm3.2\pm5.5$  &-&- \\
4.220&$41.9\pm10.7\pm3.8\pm6.6$  &-&- \\
4.230&$50.2\pm2.7\pm4.6\pm7.9$  &$25.6\pm4.8\pm2.6\pm4.0$& $0.54\pm0.11\pm0.06$ \\
4.245&$32.7\pm10.3\pm3.0\pm5.1$  &-&- \\
4.260&$41.0\pm2.8\pm3.7\pm6.4$  &$24.4\pm5.2\pm3.2\pm3.8$& $0.63\pm0.14\pm0.10$\\
4.310&$41.0\pm2.8\pm3.7\pm6.4$  &-&-\\
4.360&$52.8\pm3.7\pm4.8\pm8.2$  &$36.2\pm6.5\pm4.1\pm5.7$& $0.73\pm0.14\pm0.10$\\
4.390&$41.8\pm10.8\pm3.8\pm6.6$  &-&-\\
4.420&$49.4\pm12.4\pm4.5\pm7.6$  &-&-\\
\hline\hline
\end{tabular}
\end{center}
\end{table*}

\section{Measurement of $ e^+ e^-  \to \pi^{+}\pi^{-} h_{c}$ and observation of  $e^+ e^-  \to \pi^{0}\pi^{0} h_{c}$} \label{sec::detector}

The process of $ e^+ e^-  \to \pi^{+}\pi^{-}h_{c}$ was first observed by CLEO-c experiment using 586 $pb^{-1}$ data sample at center-of-mass (CM) energy $\sqrt{s}$ = 4.170 GeV. Together with the scan sample around 4.000 GeV and 4.260 GeV, the cross section shows a hint of rising at 4.260 GeV~\cite{lab25}, as shown in
Figure~\ref{fig fig1}. An improved measurement may shed light on understanding the nature of the Y states and also used to exam possible charged intermediate charmoniumlike states.

The BESIII experiment, works at the Beijing Electron Positron Collider (BEPC) II, is a tau-charm factory which can collect data from $\sqrt{s}$ = 2 GeV to 4.6 GeV ~\cite{lab24}. Starting from the end of 2012 to 2013, BESIII detector collected about 3.3 fb$^{-1}$ data samples at 13 CM energies from 3.900 to 4.420 GeV, which are listed in Table I.

Based on these data samples, the process of $ e^+ e^-  \to \pi\pi h_{c}$ has been studied using full reconstruction of the final state particles. Here the $h_{c}$ is reconstructed via its electric-dipole (E1) transition $h_{c}\to\gamma\eta_{c}$ with $\eta_{c}$ reconstructed with its 16 hadronic decay modes:
$\ppb$, $\ppkk$, $\pp\ppb$, $2(\kk)$, $2(\pp)$,
$3(\pp)$, $2(\pp)\kk$, $K_SK^{\pm}\pi^{\mp}$,
$K_SK^{\pm}\pi^{\mp}\pp$, $\kk\pi^0$, $\kk\eta$, $\ppb\pi^0$, $\pp\eta$,
$\pp\pi^0\pi^0$, $2(\pp)\eta$, and $2(\ppp)$. Here $K^0_S$
is reconstructed in its $\pp$ decays, and $\eta$ in
its $\gamma\gamma$ final state.

\begin{figure*}[!htbp]
\centering
\subfigure[]{ \label{fig_fig2:subfig:a}

\includegraphics[width=0.45\textwidth]{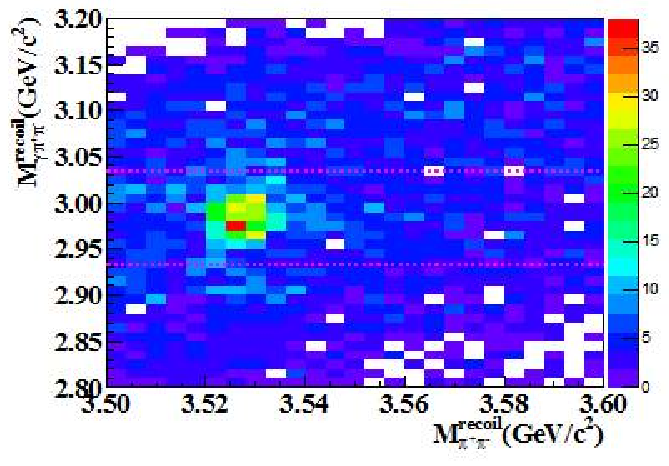}
}
\subfigure[]{ \label{fig_fig2:mini:subfig:b}
\includegraphics[width=0.43\textwidth]{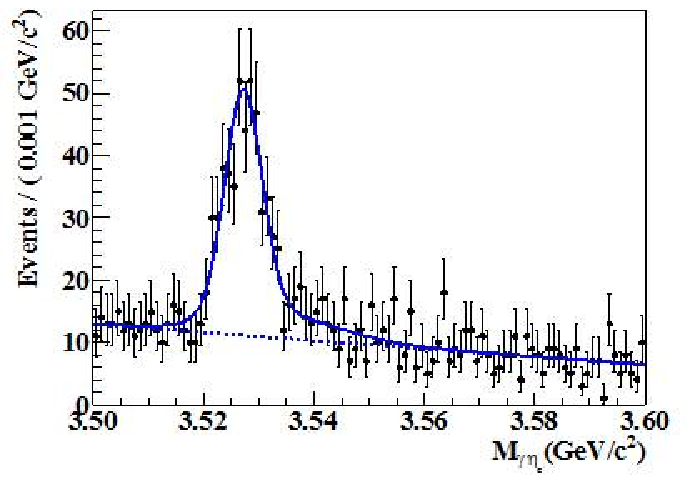}
}
\caption{ The scatter plot of the mass of the $\eta_{c}$ candidate versus that of the $h_{c}$ candidate at the CM energy of 4.260 GeV (a) and The projection of the invariant mass distribution of $\gamma\eta_{c}$ in the $\eta_{c}$ signal region, the dots with error bars are data, and the solid curve shows the best fit result (b). } \label{fig_fig2}
\end{figure*}

\begin{figure*}[!htbp]
\centering
\subfigure[]{ \label{fig_fig3:subfig:a}

\includegraphics[width=0.43\textwidth]{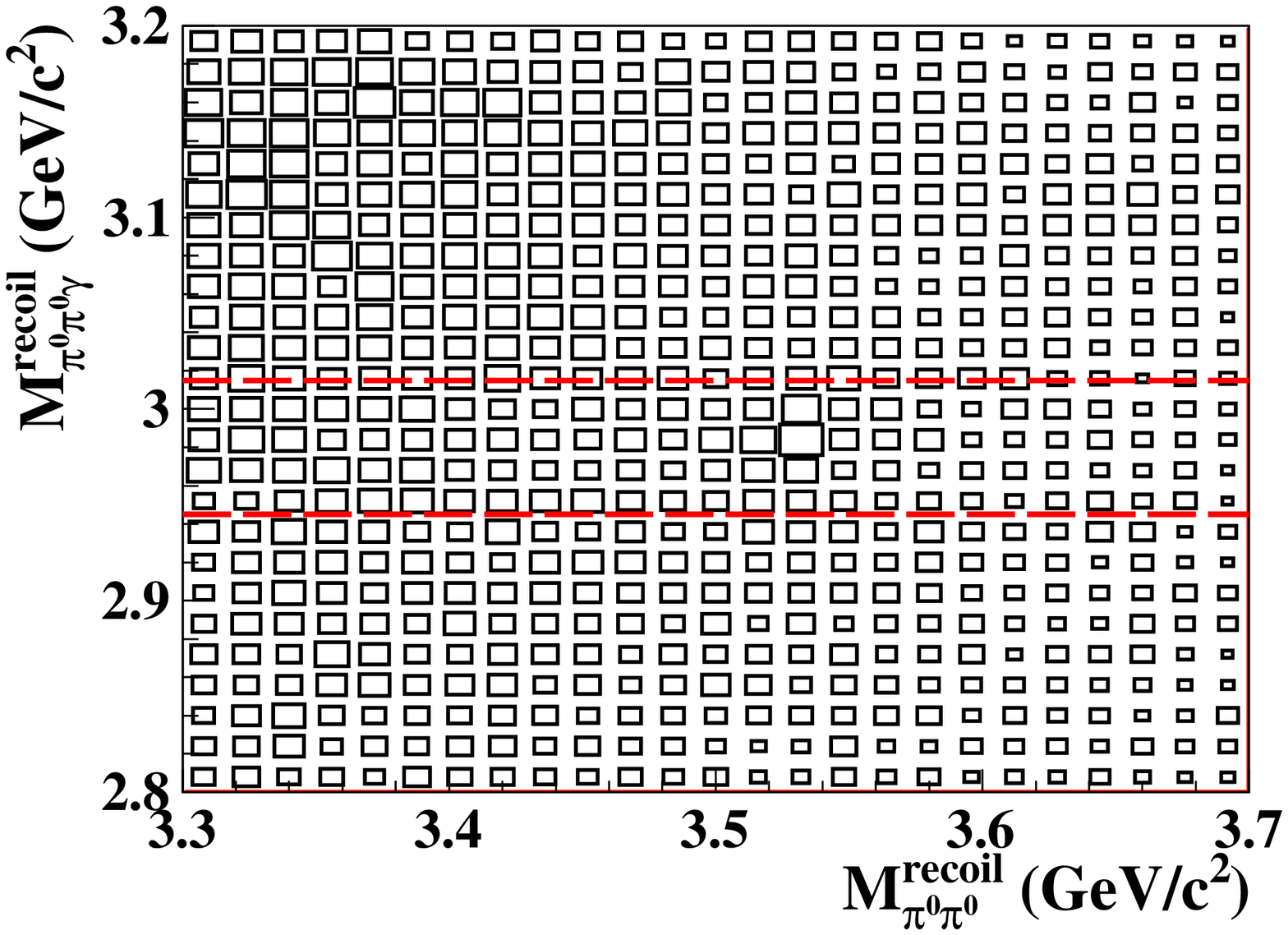}
}
\subfigure[]{ \label{fig_fig3:mini:subfig:b}
\includegraphics[width=0.43\textwidth]{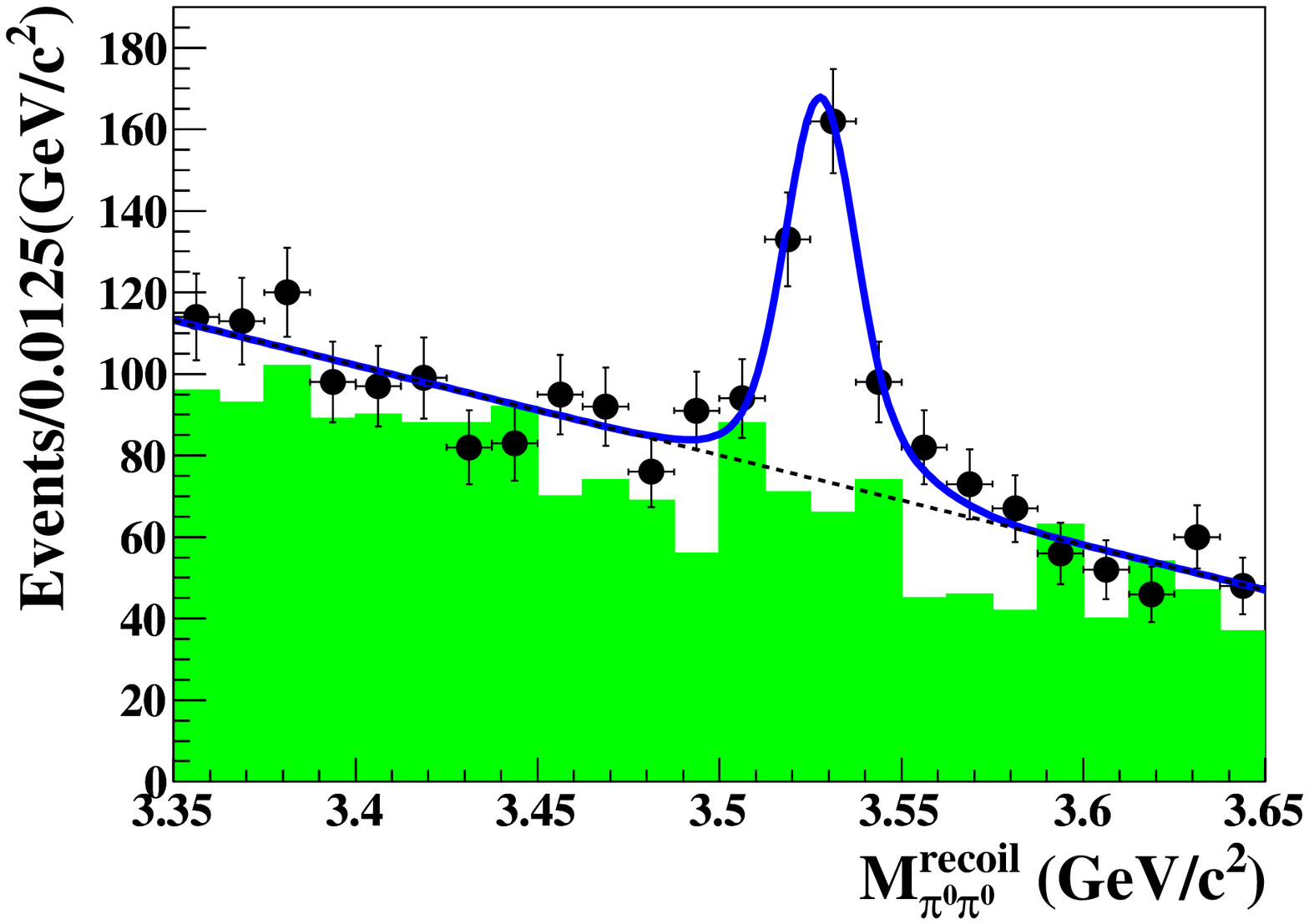}
}
\caption{ The scatter plot of the mass of the $\eta_{c}$ candidate versus that of the $h_{c}$ candidate (a) and the $M^{\rm recoil}_{\pi^{0}\pi^{0}}$ distribution for the events with an  $\eta_{c}$ candidate(b). }
\label{fig_fig3}
\end{figure*}

Figure~\ref{fig_fig2}(a) shows as an example the scatter plot of the mass of the $\eta_{c}$ candidate versus that of the $h_{c}$ candidate at the CM energy of 4.260 GeV (left panel), as well as the projection of the invariant mass distribution of $\gamma\eta_{c}$ in the $\eta_{c}$ signal region (right panel) in  $ e^+ e^-  \to \pi^{+}\pi^{-} h_{c}$ process, where a clear signal is observed~\cite{lab21}. The $\gamma\eta_{c}$ mass spectrum is used to extract the number of $\pi^{+}\pi^{-}h_{c}$ signal events by performing a fit. The fit to the 4.260 GeV data is also shown in Figure~\ref{fig_fig2}. At the energy points with large statistics (4.230, 4.260, and 4.360 GeV), the fit is applied to the 16 $\eta_{c}$ decay modes simultaneously, while at other energy points, the fit is performed to the mass spectrum summed over all the $\eta_{c}$ decay modes. With the fitted number of hc events and the corresponding detect efficiency as well as the integrated luminosity at each energy point, the cross section of  $ e^+ e^-  \to \pi^{+}\pi^{-} h_{c}$
at different energy points are listed in Table~\ref{tab::tab1}.

The cross section of $ e^+ e^-  \to \pi^{0}\pi^{0} h_{c}$  is also studied using 3 data samples at CM energy of 4.23, 4.26 and 4.36 GeV. Figure~\ref{fig_fig3}(a)
shows the scatter plot of the $\eta_{c}$ candidate versus that of the $h_{c}$ candidate summed over 3 energy points, and Figure~\ref{fig_fig3}(b) shows the projection of the invariant mass distribution of $\gamma\eta_{c}$ in the $\eta_{c}$ signal region, clear $h_{c}$ signal is also observed in the neutral process. The $\gamma\eta_{c}$ mass spectrum summed over all the $\eta_{c}$ decay modes is fitted simultaneously at the 3 energy points to extract the number of $\pi^{0}\pi^{0} h_{c}$ signal events. The cross sections of $ e^+ e^- \to \pi^{0}\pi^{0} h_{c}$ are also listed in Table~\ref{tab::tab1}.

The comparison of the cross section of  $ e^+ e^-  \to \pi^{+}\pi^{-} h_{c}$  measured at BESIII with the cross section of  $ e^+ e^-  \to \pi^{+}\pi^{-} J/\psi$  measured at Belle experiment using initial state radiative (ISR) method~\cite{lab14} is shown in Figure\ref{fig fig1}. The cross section of $ e^+ e^-  \to \pi^{0}\pi^{0} h_{c}$ is at the same order of magnitude of $\pi^{+}\pi^{-} J/\psi$ but with a different line shape. There is a possible broad structure at high energy with a local maximum around 4.23 GeV. Since the Y(4260) was established from the cross section of $ e^+ e^-  \to \pi^{+}\pi^{-} J/\psi$, the
different line shape $ e^+ e^-  \to \pi^{+}\pi^{-} h_{c}$ makes the understanding of Y states more complicate.

Combining the Born cross sections of $e^{+} e^{-} \to \pi^0\pi^0 h_{c}$ with that of charged mode, the ratios of cross section between charged and neutral mode ($R_{\pi\pi h_{c}}$) at each energy points are calculated, as listed in  Table~\ref{tab::tab1}. In the calculation, the common systematic uncertainties in the two measurements have been canceled. The combined ratio for the 3 energy points $R_{\pi\pi h_{c}}$ , obtained with a weighted least squares method~\cite{lab27}, is determined to be $0.63\pm0.09$, which agrees with the expectation based on isospin symmetry within a systematic uncertainty of 0.5.

 \begin{figure*}[!htbp]
\centering
\subfigure[]{ \label{fig fig4:subfig:a}

\includegraphics[width=0.43\textwidth]{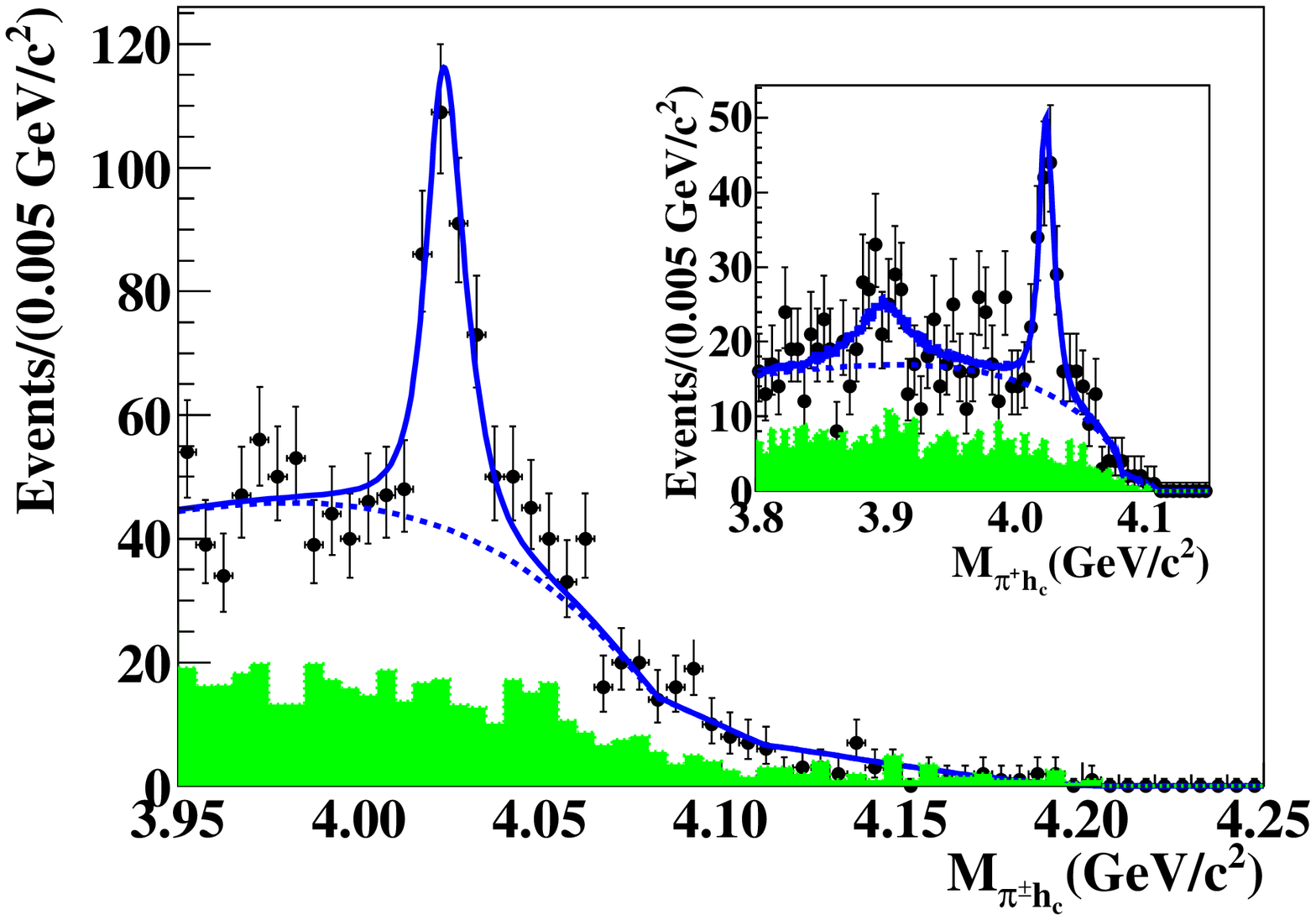}
}
\subfigure[]{ \label{fig fig4:mini:subfig:b}
\includegraphics[width=0.43\textwidth]{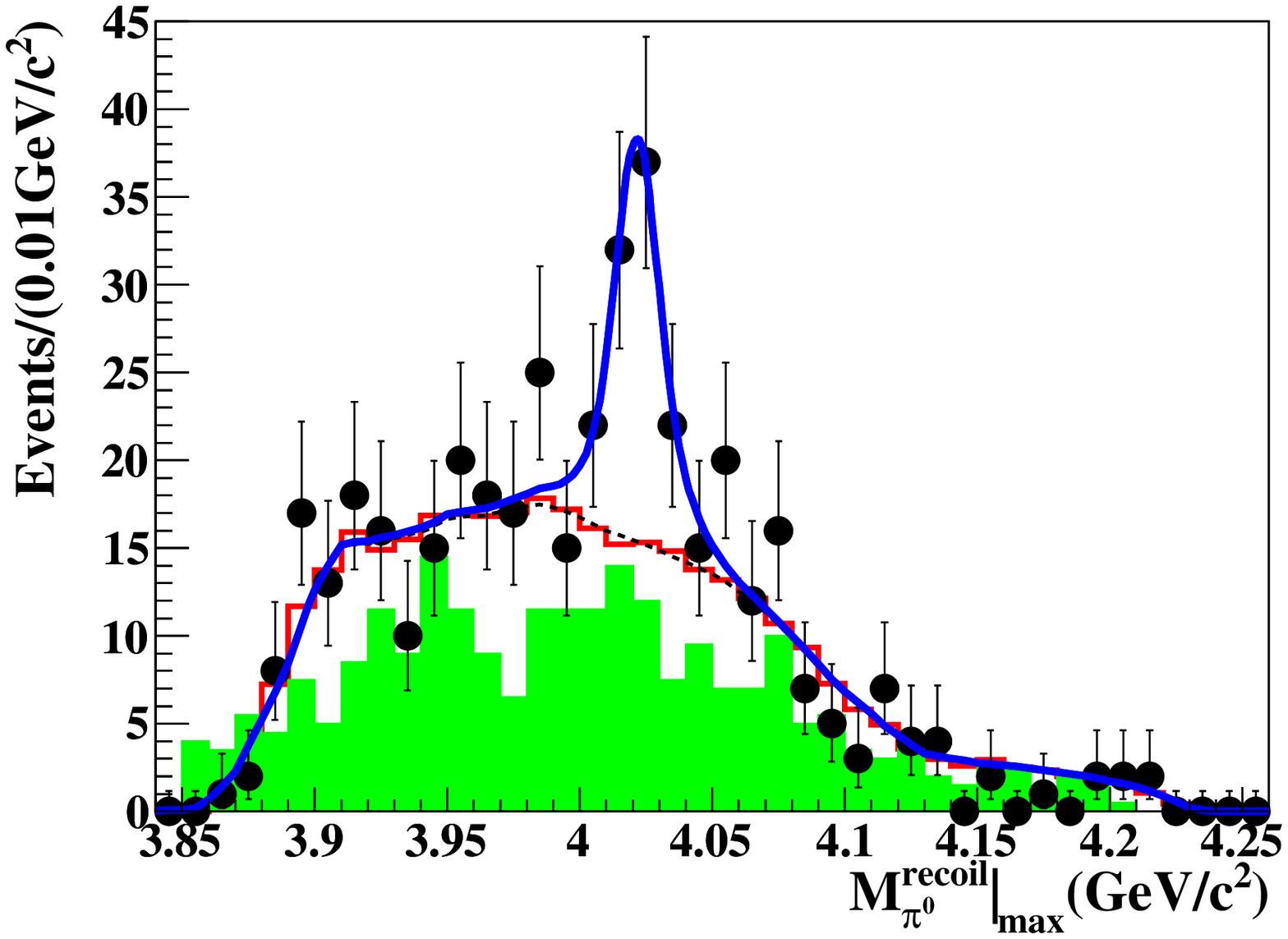}
}
\caption{ Sum of the simultaneous fit to the $M_{\pi^{\pm}h_{c}}$ distribution (a) and sum of the simultaneous fit to the $M^{\rm recoil}_{\pi^{0}}|_{\rm max}$ distribution at $\sqrt{s} = $ 4.23, 4.26 and 4.36 GeV as described in the text(b). } \label{fig fig4}
\end{figure*}

\begin{table*}[!htbp]
\begin{center}
\renewcommand{\arraystretch}{1.2}
\caption{ Cross sections of $e^+ e^-\to \pi Z_{c}(4020)\to\pi\pi h_{c}$ measured from the BESIII experiment and ratios $R_{\pi Z_{c}(4020)}$ mentioned in the text. The first errors are statistical errors, the second errors are systematic. Here, $\sigma^{\rm B}_{\pi^{\mp}Z_{c}(4020)^{\pm}}$ denotes $\sigma^{\rm B}(e^+ e^-\to \pi^{\mp} Z_{c}(4020)^{\pm}\to\pi^+\pi^- h_{c})$ and $\sigma^{\rm B}_{\pi^{0}Z_{c}(4020)^{0}}$ denotes $\sigma^{\rm B}(e^+ e^-\to \pi^{0} Z_{c}(4020)^{0}\to\pi^0\pi^0 h_{c})$.  } \label{tab::tab2}
\begin{tabular}{c c c c c c}
\hline \hline
$\sqrt{s}$(GeV) & $n_{Z_{c}(4020)^{\pm}}^{\rm obs}$ & {$\sigma^{\rm B}_{\pi^{\mp}Z_{c}(4020)^{\pm}}$($\rm pb$)}
&$n_{Z_{c}(4020)^{0}}^{\rm obs}$  & {$\sigma^{\rm B}_{\pi^{0}Z_{c}(4020)^{0}}$($\rm pb$)}& $R_{\pi Z_{c}(4020)}$\\\hline
4.23&$114\pm25$ & $8.4\pm1.8\pm2.6$ &$22\pm7$& $6.5\pm2.2\pm0.7$ & $0.77\pm0.31\pm0.25$ \\
4.26&$76\pm17$ & $7.0\pm1.6\pm2.0$ &$23\pm8$& $8.5\pm2.9\pm1.1$ & $1.21\pm0.50\pm0.38$ \\
4.36&$67\pm15$ & $9.8\pm2.2\pm2.9$ &$17\pm7$ & $9.9\pm4.1\pm1.3$ & $1.00\pm0.48\pm0.32$ \\
\hline\hline
\end{tabular}
\end{center}
\end{table*}
%
\section{Observation of the charmoniumlike state $Z_{c}(4020)$} \label{sec::selection}
At the three energy points with large statistics, the intermediate states are studied by examining the Dalitz plot of the selected hc candidate events.
 Figure~\ref{fig fig4}(a) shows the projection of the $M(\pi^{\pm}h_{c})$ (two entries per event) distribution for the $\pi^{+}\pi^{-}h_{c}$ candidate events summed over 3 energy points. There is a significant peak at around 4.02 GeV/c$^{2}$ (named $Z_{c}(4020)^{\pm}$ hereafter), and there are also some events around 3.9 GeV/c$^{2}$ (inset of Figure~\ref{fig fig4}(a)), which may be the candidate events of the observed $Z_{c}(3900)^{\pm}$ in $e^{+}e^{-}\to\pi^{+}\pi^{-}J/\psi$~\cite{lab13}. In the $\pi\pi$ mass spectrum, there is no obvious structure.

Similarly, in the  $M^{\rm recoil}_{\pi^{0}}|_{\rm max}$ (one entry per event, the larger one in two $M^{rec}_{\pi^{0}}$ candidates is retained) mass distribution for the $e^{+} e^{-} \to \pi^{0}\pi^{0}h_{c}$ events, there is also a significant peak, which could be the isospin partner of $Z_{c}(4020)^{\pm}$,  at around 4.02 GeV/c$^{2}$(named $Z_{c}(4020)^{0}$ hereafter).

$M(\pi^{\pm}h_{c})$ or $M^{\rm recoil}_{\pi^{0}}|_{\rm max}$ distributions summed over the 16 $\eta_{c}$ decay modes is fitted simultaneously at the three energy points to extract the parameters of the structures and the signal events yield.The mass and width of $Z_{c}(4020)^{\pm}$ are measured to be
$4023.6 \pm 2.2 \pm 3.8$ MeV/$c^2$, and $7.9 \pm 2.7 \pm 2.6$ MeV, respectively, with statistical significance of 8.9¦Ò (see Figure~\ref{fig fig4}(a)), while the mass of $Z_{c}(4020)^{0}$ is determined to be ($4023.6 \pm 2.2 \pm 3.8$) MeV/$c^2$ with a statistical significance larger than 5$\sigma$ (see Figure~\ref{fig fig4}(b)).
Here, the width of $Z_{c}(4020)^{0}$ is fixed as its charged isospin partner due to the limitation of signal events. The detail information about the systematic uncertainties in mass and width measurement of $Z_{c}(4020)$ can be found in published paper~\cite{lab21,lab22}. In the charged mode, the possible decay of $Z_{c}(3900)^{\pm}\to\pi^{\pm}h_{c}$ has been explored by adding a $Z_{c}(3900)$ structure in the fit with mass and width fixed to the BESIII measurement~\cite{lab9}. It is found the statistical significance of $Z_{c}(3900)$ is less than 2.1$\sigma$ (see the inset of Figure~\ref{fig fig4}(a))~\cite{lab20}.

The signal yields of $Z_{c}(4020)$, as well as the measured cross sections for $e^{+}e^{-} \to \pi Z_{c}(4020) \to \pi\pi h_c$ and the ratios ${\cal R}_{\pi Z_{c}}$ between neutral and charged modes at three energy points, are listed in Table~\ref{tab::tab2}. The common systematic uncertainty in the ratio calculation has been canceled. The combined ratio ${\cal R}_{\pi Z_{c}}$ for 3 energy points  is determined to be $(0.99 \pm 0.31)$ with the same method as for the combined ${\cal R}_{\pi\pi h_{c}}$, which is consistent well with the expectation of isospin symmetry = 0.5 within 1.6$\sigma$.

\begin{table}[!htbp]
\begin{center}
\renewcommand{\arraystretch}{1.2}
\caption{ The resonance parameters of the $Z_{c}(4020)$ and $Z_{c}(4025)$.} \label{tab::tab3}
\begin{tabular}{c c c }
\hline \hline
State & Mass (MeV/c$^{2}$) & Width (MeV)\\\hline
$Z_{c}(4020)^{\pm}$&$4022.9\pm0.8\pm2.7$ & $7.9\pm2.7\pm2.6$  \\
$Z_{c}(4020)^{0}$&$4023.9\pm2.2\pm3.8$ & Fixed (=7.9 MeV) \\
$Z_{c}(4025)^{\pm}$&$4026.3\pm2.6\pm3.7$ & $24.6\pm5.6\pm3.7$  \\
\hline\hline
\end{tabular}
\end{center}
\end{table}

\section{Summary And Outlook}
In summary, the charmoniumlike structure, $Z_{c}(4020)$, has been observed in $\pi h_{c}$ system for both charged and neutral modes. It is not only a multi-quark state, but also a isospin triplet. Table~\ref{tab::tab3} shows the measured resonance parameters of the charged and neutral $Z_{c}(4020)$ state.
The mass difference between neutral and charged $Z_{c}(4020)$ is $1.0\pm2.2$ (stat.) MeV/c$^{2}$,
which in good agreement with zero within uncertainty. Furthermore, the measured Born cross sections for $\sigma(e^{+} e^{-} \to \pi^{0}\pi^{0}h_c)$ are about half of those for  $\sigma(e^{+} e^{-} \to \pi^{+}\pi^{-}h_c)$ within errors, while the measured Born cross sections $\sigma(e^{+} e^{-} \to \pi^{0}\pi^{0}h_c)$  for charged and neutral modes.

Furthermore, the measured Born cross sections for $\sigma(e^{+} e^{-} \to \pi^{0}\pi^{0}h_c)$ are about half of those for $\sigma(e^{+} e^{-} \to \pi^{+}\pi^{¨C}h_c)$ within errors, while the measured Born cross sections $\sigma(e^{+} e^{-} \to \pi Z_c(4020))$ for charged and neutral modes are generally consistent. It indicates that there is no obvious isospin violations in $\pi\pi h_{c}$ and $\pi Z_{c}(4020)$ systems.

The observation of charged $Z_{c}(4020)$ confirms that there is an extra multi-quark state except the observed $Z_{c}(3900)$. In particular, c is the first observed neutral charmoniumlike Z state, and it is helpful to understanding the nature of $Z_{c}$ and even other XYZ particles. For instance, the observation of theoretical expectations of $Z_{c}(4020)$ and $X(3872)$ combined with $Y(4260)$ transition to $X(3872)$, which establish contact with XYZ particles, issue the challenge to interpret these particles theoretically.

 Compared to the $Z_{c}(4025)$ observed in the process $e^{+}e^{-}\to (D^{*}\bar{D^{*}})^{\pm}\pi^{\mp}$~\cite{lab23}, the resonance parameters of $Z_{c}(4020)$~\cite{lab21,lab22} agree with $Z_{c}(4025)$~\cite{lab23} with 1.5$\sigma$. If they are the same state, the coupling of $Z_{c}(4020)$
 to $D^{*}D^{*}$ final state is about 12 times larger than to $\pi h_{c}$ final state. To clarify whether they are the same particle, one needs a further sophisticated analysis with a coupled channel technique.

 The observation of charged charmonium-like family members together with the similar observation in bottomonium family provides important evidence for the existence of hadrons beyond the conventional quark model and QCD theory, and helps in the improvement of our understanding about the fundamental structure of the matter.

In the near future, more data between 4.0 and 4.6 GeV at BESIII experiment~\cite{lab24} will be accumulated for further study of XYZ states. The property of $Z_{c}(4020)$ will be studied more carefully. Meantime, the line-shape of cross section of $e^{+}e^{-}\to\pi\pi h_{c}$ and $e^{+}e^{-}\to\pi Z_{c}(4020)$ will be investigated, which can shed light on the understanding of this state and even other XYZ particles,

{\it This work is supported in part by Joint Large-Scale Scientific Facility Funds of the NSFC and CAS under Contracts No. U1332201, and National Natural Science Foundation of China (NSFC) under Contracts No. 11475207.}


\end{document}